\documentclass[11pt,a4paper,english]{amsart}
\usepackage[utf8]{inputenc}
\usepackage[english]{babel} 

\usepackage{amsmath} %Numberwithin
\usepackage{amsthm} %Cambiar lo de teorema, proposición...
\usepackage{graphicx} 
\usepackage{xcolor} %Imprimir palabras de otro color
\usepackage{amsfonts}
\usepackage[biblabel]{cite}
\usepackage{bm} %Letra en negrita en math mode
\usepackage[hidelinks]{hyperref} %Poner links
\usepackage{amssymb} %Usar \supsetneq
\usepackage{tikz-cd} %Diagramas
\usepackage{amscd}
\usepackage{etoolbox}% http://ctan.org/pkg/etoolbox
\patchcmd{\thmhead}{(#3)}{#3}{}{}

\DeclareMathOperator{\ev}{ev} %Aplicación de evaluación
\DeclareMathOperator{\PRM}{PRM}
\DeclareMathOperator{\RM}{RM}

\DeclareMathOperator{\wt}{wt}
\DeclareMathOperator{\Hull}{Hull}

\newcommand{\F}{{\mathbb{F}}}
\newcommand{\fq}{\mathbb{F}_q}
\newcommand{\fqs}{\mathbb{F}_{q^s}}

\newcommand{\PM}{{\mathbb{P}^{m}}}

\usepackage{mathtools} %Valor abs y norma. Con * no se ajusta la altura.
\DeclarePairedDelimiter\abs{\lvert}{\rvert}%
\DeclarePairedDelimiter\norm{\lVert}{\rVert}%

\makeatletter
\let\oldabs\abs
\def\abs{\@ifstar{\oldabs}{\oldabs*}}
\let\oldnorm\norm
\def\norm{\@ifstar{\oldnorm}{\oldnorm*}}
\makeatother

\newtheorem{thm}{Theorem}[section]
\newtheorem{prop}[thm]{Proposition}

\newtheorem{lem}[thm]{Lemma}
\theoremstyle{definition}

\newtheorem{rem}[thm]{Remark} 
 
\newtheorem{ex}[thm]{Example}

\title[EAQECCs from projective Reed-Muller codes and hull variation]{Quantum error-correcting codes from projective Reed-Muller codes and their hull variation problem}
\author{Diego Ruano and Rodrigo San-José}
\curraddr{
\texttt{Diego Ruano, Rodrigo San-José:} IMUVA-Mathematics Research Institute, Universidad de Valladolid, 47011 Valladolid (Spain).
}
\email{diego.ruano@uva.es; rodrigo.san-jose@uva.es}

\thanks{This work was supported in part by the following grants: Grant TED2021-130358B-I00 funded by MCIN/AEI/10.13039/501100011033 and by the ``European Union NextGenerationEU/PRTR'', by Grant PID2022-138906NB-C21 funded by MICIU/AEI/10.13039/501100011033 and by ERDF/EU, QCAYLE project funded by MCIN, the European Union NextGenerationEU (PRTR C17.I1) and Junta de Castilla y Le\'on, and FPU20/01311 funded by the Spanish Ministry of Universities.}



\begin{document}

\maketitle

\begin{abstract}
Long quantum codes using projective Reed-Muller codes are constructed. Projective Reed-Muller codes are evaluation codes obtained by evaluating homogeneous polynomials at the projective space. We obtain asymmetric and symmetric quantum codes by using the CSS construction and the Hermitian construction, respectively. We provide entanglement-assisted quantum error-correcting codes from projective Reed-Muller codes with flexible amounts of entanglement by considering equivalent codes. Moreover, we also construct quantum codes from subfield subcodes of projective Reed-Muller codes.
\end{abstract}

\begin{flushright}
{\em Dedicated to Professor Sudhir R. Ghorpade on the occasion of his sixtieth birthday.}
\end{flushright}

\section{Introduction}

Stabilizer quantum error-correcting codes (QECCs) can be defined from classical linear codes. They can be defined from a pair of self-orthogonal classical linear codes with respect to the Euclidean inner product, CSS construction, and from a self-orthogonal classical linear code with respect to the Hermitian inner product, Hermitian construction    \cite{cssoriginal1,cssoriginal2,kkks}. Moreover, by sharing entanglement between encoder and decoder, it is possible to increase the communication
capacity and remove the self-orthogonality condition, giving rise to entanglement-assisted quantum error-correcting codes (EAQECCs) \cite{entanglement}. For the CSS-like construction, the minimum number of maximally entangled quantum states required is equal to $c:=\dim C_1-\dim C_1\cap C_2^\perp$. Therefore, the dimension of the \textit{relative hull of $C_1$ with respect to $C_2$}, which is defined in \cite{relativehull} as $\text{Hull}_{C_2}(C_1):=C_1\cap C_2^\perp$, determines the parameter $c$. For the Hermitian construction, we only use one code $C$, and the parameter $c$ is given by $\dim C-\dim C\cap C^{\perp_h}$, where $C^{\perp_h}$ is the Hermitian dual of $C$. The \textit{Hermitian hull} of $C$ is thus defined as $\text{Hull}_H(C):=C\cap C^{\perp_h}$. Moreover, both QECCs and EAQECCs can be considered in an asymmetric fashion taking advantage of the asymmetry in quantum
errors since phase-shift errors are more probable than qudit-flip errors \cite{ioffe,asymmetricConstructions,galindoasymmetric}. 

In this paper, we construct QECCs and EAQECCs from projective Reed-Muller codes, which is a family of evaluation codes obtained by evaluating homogeneous polynomials of a given degree at the projective space \cite{lachaud,sorensen}. In \cite{lachaud}, it is shown that these codes can outperform affine Reed-Muller codes in some instances. In general, one needs to require entanglement assistance when working with evaluation codes over the projective space, in particular, because the resulting families of codes are not necessarily nested. EAQECCs from projective Reed-Muller codes have been studied in \cite{sanjoseSSCPRS,sanjoseSSCPRM,sanjoseHullsPRM}, and in \cite{libroprm} for the few cases in which one has the nested condition. In Section \ref{seceu}, we deal with asymmetric quantum codes coming from the CSS construction, and, in Section \ref{secherm}, we deal with (symmetric) quantum codes coming from the Hermitian construction. 

More concretely, we provide EAQECCs with a flexible amount of entanglement and unassisted EAQECCs, that is, the parameter $c$ is equal to zero, which simply corresponds to the case of QECCs. We achieve this by considering equivalent linear codes whose (relative) hull has different dimension, i.e., we consider the so-called hull variation problem for projective Reed-Muller codes (following the terminology from \cite{haochenhull}). We remark that it is always possible to increase the parameter $c$ one by one \cite{relativehull,grasslvariableentanglement}, for $q>2$, however, one can only decrease the parameter $c$ under certain conditions \cite{relativehull,haochenhull}, for $q>2$. Note that, by considering equivalent codes that give rise to a different parameter $c$, the rate of the quantum code varies but the net rate is preserved (see Remark \ref{rem:rate} for details). Having a rich family of codes with the same net rate but with a different minimum number of entangled quantum states provides flexibility for practical applications. Moreover, the unassisted case is especially interesting, $c=0$, since it has a simpler implementation. This corresponds to the case where the dimension of the relative hull is equal to $\dim C_1$ or the dimension of the Hermitian hull is equal to $\dim C$.

For both constructions, in Sections \ref{seceu} and \ref{secherm}, we find conditions to obtain unassisted QECC by using projective Reed-Muller codes, obtaining asymmetric and symmetric quantum codes with good parameters, they surpass the Gilbert-Varhsamov bound \cite{quantumGVexigente,matsumotoimprovedGV}. Moreover, we show that the quantum codes obtained outperform the ones obtained from affine Reed-Muller codes. On top of this, we consider quantum codes from the subfield subcodes of projective Reed-Muller codes. That is, given $C\subset \F_{q^s}^n$, we consider $C_q:=C\cap \fq^n$, its \textit{subfield subcode} with respect to the extension $\fqs/\fq$. We remark that this approach is not usually fruitful for this family of codes since one does not have conditions for having nested codes. Nevertheless, we are able to find certain technical conditions that allow us to consider subfield subcodes and construct long quantum codes over small finite fields.  

\section{Preliminaries}
Let $\fq$ be the finite field with $q$ elements. We denote by $\PM$ the projective space over $\fq$. We choose for $\PM$ the standard representatives, i.e., the representatives whose leftmost nonzero coordinate is equal to 1. If we regard the standard representatives as points in the affine space $\mathbb{A}^{m+1}$, we obtain the set $P^m:=\{Q_1,\dots,Q_n\}\subset\mathbb{A}^{m+1}$, where $n=\abs{P^m}=\frac{q^{m+1}-1}{q-1}$. 

Let $S=\fq [x_0,\dots,x_m]$. Given $d$ a positive integer, we denote $S_d\subset S$ the set of homogeneous polynomials of degree $d$. We define the evaluation map
$$
\ev:S_d \rightarrow \fq^{n},\:\: f\mapsto \left(f(Q_1),\dots,f(Q_n)\right).
$$
The image of $\ev$ is the \textit{projective Reed-Muller code of degree} $d$, denoted by $\PRM_d(q,m)$, or $\PRM_d(m)$ if there is no confusion about the field. For the minimum distance of a code $C\subset \fq^n$, we use the notation $\wt(C)$. We have the following results about the parameters of projective Reed-Muller codes and their duality from \cite{sorensen} (also see \cite{ghorpadeMinimumweightPRM}). 

\begin{thm}\label{paramPRM}
The projective Reed-Muller code $\PRM_d(q,m)$, $1\leq d\leq m(q-1)$, is an $[n,k]$-code with 
$$
\begin{aligned}
&n=\frac{q^{m+1}-1}{q-1},\\
&k=\sum_{t\equiv d\bmod q-1,0<t\leq d}\left( \sum_{j=0}^{m+1}(-1)^j\binom{m+1}{j}\binom{t-jq+m}{t-jq}   \right).\\
\end{aligned}
$$
For the minimum distance, we have
$$
\wt(\PRM_d(q,m))=(q-\ell)q^{m-r-1}, \text{ where } \;
d-1=r(q-1)+\ell, \;0\leq \ell <q-1.
$$
\end{thm}
\begin{thm}\label{dualPRM}
Let $1\leq d\leq m(q-1)$ and let $d^\perp=m(q-1)-d$. Then
$$
\begin{aligned}
&\PRM_d^\perp(q,m)=\PRM_{d^\perp}(q,m) &\text{ if } d\not\equiv 0\bmod (q-1), \\
&\PRM_d^\perp(q,m)=\PRM_{d^\perp}(q,m)+\langle (1,\dots,1) \rangle &\text{ if } d\equiv 0\bmod (q-1).
\end{aligned}
$$
\end{thm}

With respect to affine Reed-Muller codes, we will denote them by $\RM_d(q,m)$, or by $\RM_d(m)$ if there is no confusion about the field. We recall now the analogous results for affine Reed-Muller codes \cite{delsarteRM,kasamiRM}.

\begin{thm}\label{paramRM}
The Reed-Muller code $\RM_d(q,m)$, $1\leq d\leq m(q-1)$, is an $[n,k]$-code with 
$$
\begin{aligned}
&n=q^m,\\
&k=\sum_{t=0}^d \sum_{j=0}^m (-1)^j\binom{m}{j}\binom{t-jq+m-1}{t-jq}.\\
\end{aligned}
$$
For the minimum distance, we have
$$
\wt(\RM_d(q,m))=(q-\ell)q^{m-r-1}, \text{ where } \;
d=r(q-1)+\ell, \;0\leq \ell <q-1.
$$
\end{thm}

\begin{thm}\label{dualRM}
Let $1\leq d\leq m(q-1)$. Then
$$
\RM_d^\perp(q,m)=\RM_{m(q-1)-d-1}(q,m).
$$
\end{thm}

For comparisons between the projective case and the affine case, it is important to note that $\wt(\PRM_d(q,m))=\wt(\RM_{d-1}(q,m))$.

Regarding quantum codes, we will consider the CSS construction and the Hermitian construction. For the CSS construction, we are considering asymmetric QECCs. In quantum error-correction we may consider two different types of errors, phase-shift errors and qudit-flip errors. As the likelihood of occurrence of each type of error is different \cite{ioffe}, it is desirable to have different error correction capabilities for each type of error, which is what asymmetric QECCs accomplish. Asymmetric QECCs have two minimum distances, $\delta_z$ and $\delta_x$, meaning that they can correct up to $\lfloor(\delta_z-1)/2\rfloor$ phase-shift errors and $\lfloor(\delta_x-1)/2\rfloor$ qudit-flip errors. We denote the parameters of an asymmetric EAQECC by $[[n,\kappa,\delta_z/\delta_x;c]]_q$, where $n$ is the length, $\kappa$ is the dimension, and $c$ is the minimum number of maximally entangled quantum qudit pairs required. We state now the CSS construction for asymmetric EAQECCs \cite{galindoasymmetric}. 

\begin{thm}[(CSS Construction)]\label{asimetricos}
Let $C_i\subset \fq^n$ be linear codes of dimension $k_i$, for $i=1,2$. Then, there is an asymmetric EAQECC with parameters $[[n,\kappa,\delta_z/\delta_x;c]]_q$, where
$$
\begin{aligned}
c&=k_1-\dim (C_1\cap C_2^\perp), \; \kappa=n-(k_1+k_2)+c, \\
\delta_z=\wt&\left(C_1^\perp\setminus \left(C_1^\perp\cap C_2 \right)\right) \text{ and }\; \delta_x=  \wt\left(C_2^\perp\setminus \left(C_2^\perp\cap C_1 \right) \right) .
\end{aligned}
$$
\end{thm}
Let $\delta^*_z:=\wt(C_1^\perp)$ and $\delta^*_x:=\wt(C_2^\perp)$. If $\delta_z=\delta^*_z$ and $\delta_x=\delta^*_x$, we say that the corresponding EAQECC is \textit{pure}, and we say it is \textit{impure} if $\delta_z>\delta^*_z$ or $\delta_x>\delta^*_x$. 

As we stated in the introduction, we are interested in constructing QECCs with a flexible amount of entanglement and, in particular, without entanglement assistance. This corresponds to pairs of codes $C_1,C_2$ such that $\dim \text{Hull}_{C_2}(C_1)=\dim C_1$, equivalently, $C_1\subset C_2^\perp$.

For the Hermitian construction, we need to consider codes defined over $\F_{q^2}^n$ and the Hermitian product. For two vectors $v,w\in \F_{q^2}^n$, their Hermitian product is 
$$
v \cdot_h w:=\sum_{i=1}^n v_iw_i^q.
$$
The Hermitian dual of a code $C\subset \F_{q^2}^n$ is defined as $C^{\perp_h}:=\{v\in \F_{q^2}^n\mid v\cdot_h w=0, \;\forall\; w \in C \}$. 

\begin{rem}
For a code $C\subset \F_{q^2}^n$, we have that $C^{\perp_h}=(C^{\perp})^q$, where we consider the component wise power of $q$. This implies that the Euclidean dual and the Hermitian dual codes have the same parameters. 
\end{rem}

We can state now the Hermitian construction for EAQECCs \cite{entanglement,galindoentanglement}.

\begin{thm}[(Hermitian construction)]\label{hermitica}
Let $C\subset \F_{q^2}^n$ be a linear code of dimension $k$ and $C^{\perp_h}$ its Hermitian dual. Then, there is an EAQECC with parameters $[[n,\kappa,\delta;c]]_q$, where
$$
c=k-\dim(C\cap C^{\perp_h}), \;\kappa=n-2k+c, \; \text{ and } \;\delta=\wt(C^{\perp_h}\setminus (C\cap C^{\perp_h})).
$$
\end{thm}

From the Hermitian construction for EAQECCs we can only obtain symmetric EAQECCs, that is, EAQECCs with $\delta_z=\delta_x=\delta$. Thus, if we define $\delta^*:=\wt(C^{\perp_h})$, the corresponding quantum code is impure if $\delta>\delta^*$, and pure otherwise. 

For asymmetric QECCs (EAQECCs with $c=0$), we will use the notation $[[n,\kappa,\delta_z/\delta_x]]_q$ for their parameters. Analogously, for symmetric QECCs we will use $[[n,\kappa,\delta]]_q$. In all of our results, we compute $\delta_z^*$ and $\delta_x^*$ ($\delta^*$ respectively for the Hermitian construction), thus obtaining lower bounds for the true minimum distance of the corresponding EAQECCs. 

\section{CSS construction}\label{seceu}
In this section, we focus on the QECCs obtained using the CSS Construction \ref{asimetricos} with projective Reed-Muller codes. We do this by increasing the size of the relative hull via equivalent codes. We give some preliminaries first. Given two vectors $v,u\in\fq^n$, we use the following notation:
$$
u\star v:=(u_1v_1,\dots,u_nv_n) \in \fq^n.
$$
For two codes $C_1,C_2$ in $\fq^n$, we consider
$$
C_1\star C_2:=\langle u_1\star u_2 \mid u_1\in C_1, \; u_2\in C_2\rangle.
$$
We say that $C_1$ is monomially equivalent to $C_2$ if there is some vector $v\in \fq^n$ with Hamming weight $n$ such that $C_1=\langle v \rangle \star C_2$. In more generality, we will say that two codes $C$ and $C'$ are \textit{equivalent} if there exists a vector $v\in \F_q^n$ and a permutation $\sigma$ such that
$$
C=\langle v \rangle \star \sigma(C').
$$
It is clear that equivalent codes have the same basic parameters. Moreover, due to MacWilliam's Theorem \cite{MacWilliamsIsometries}, every isometry on $\F_q^n$ with respect to the Hamming metric can be obtained in this way for some vector $v\in \F_q^n$ and some permutation $\sigma$. The following result from \cite{relativehull} allows us to increase the dimension of the relative hull of projective Reed-Muller codes in some cases via equivalent codes.

\begin{thm}\label{subirhull}
For $i=1,2$, let $C_i$ be $[n,k_i]_q$ codes with $q>2$. For any $\ell$ with $\max\{0,k_1-k_2\}\leq \ell \leq \max \wt((C_1\star C_2)^\perp)-n+k_1$, there exists a code $C_{1,\ell}$ equivalent to $C_1$ such that
$$
\dim \Hull_{C_2}(C_{1,\ell})=\ell.
$$
In particular, if $\max \wt((C_1\star C_2)^\perp)=\min \{n,2n-k_1-k_2\}$, $\ell$ runs over all the possible values of $\dim \Hull_{C_2}(C_1')$, where $C_1'$ is a code equivalent to $C_1$.
\end{thm}

\begin{rem}\label{rem:rate}
For a quantum code $C$ with parameters $[[n,\kappa,\delta;c]]_q$, the rate and net rate are defined as $\rho:=\kappa/n$ and $\overline{\rho}:=(\kappa-c)/n$, respectively. If this code is constructed with two codes $C_1,C_2$ using Theorem \ref{asimetricos}, and we consider an equivalent code $C_{1,\ell}$ as in Theorem \ref{subirhull}, then the resulting quantum code has the same net rate. The same happens with codes constructed with the Hermitian construction.
\end{rem}

The following lemma is key to study the orthogonality and will be used in the subsequent results. 

\begin{lem}\label{lemasumafq}
Let $\gamma$ be a non-negative integer, and $x^\gamma\in \fq[x]$. We have the following:
$$
\sum_{z\in\F_q}x^\gamma(z)=
\begin{cases}
0 &\text{ if } \gamma=0 \text{ or } \gamma>0 \text{ and } \gamma\not\equiv 0 \bmod (q-1),\\
-1 &\text{ if } \gamma>0 \text{ and } \gamma\equiv 0 \bmod (q-1).
\end{cases}
$$
\end{lem}
\begin{proof}
Let $\xi\in \F_{q}$ be a primitive element. Then $\fq=\{\xi^0,\xi^1,\dots,\xi^{N-2}\}\cup\{0\}$. If $\gamma=0$, $x^\gamma=1$, and the sum is equal to $\abs{\fq}=q=0$ in $\fq$. If $\gamma>0$ and $\gamma\equiv 0 \bmod (q-1)$, then $x^\gamma(z)=1$ for all $z\in \fq^*$, and $\sum_{z\in \fq}x^\gamma(z)=q-1=-1$. Finally, if $\gamma>0$ and $\gamma\not \equiv 0 \bmod (q-1)$, we have
$$
\sum_{z\in \fq}x^\gamma(z)=\sum_{i=0}^{q-2}(\xi^{i})^\gamma=\frac{\xi^{\gamma(q-1)}-1}{\xi^\gamma -1}=0.
$$
\end{proof}

\begin{rem}\label{remafin}
When working over the affine space $\mathbb{A}^\ell$, for $1\leq \ell \leq m$, if we consider $x_1^{\alpha_1}\cdots x_\ell^{\alpha_\ell}\in \fq[x_1,\dots,x_m]$, we have 
$$
\sum_{Q\in \mathbb{A}^\ell} x_1^{\alpha_1}\cdots x_\ell^{\alpha_\ell}(Q)=\left( \sum_{z\in \fq}x_1^{\alpha_1}(z) \right)\cdots \left( \sum_{z\in \fq}x_\ell^{\alpha_\ell}(z) \right).
$$
Thus, we can use Lemma \ref{lemasumafq} in order to obtain the result of this sum. In particular, if we have $\alpha_i<q-1$ for some $1\leq i \leq \ell$, this sum is equal to 0.
\end{rem}

To increase the dimension of the hull using Theorem \ref{subirhull}, we first note that, if $C_i=\PRM_{d_i}(m)$, for $i=1,2$, then $C_1\star C_2=\PRM_{d_1+d_2}(m)$. If $k_1+k_2>n$, from \cite[Lem. 4.8]{randriamSchurproduct} we obtain that $\wt(C_1\star C_2)=\wt(\PRM_{d_1+d_2}(m))=1$. By Theorem \ref{paramPRM}, this implies that $d_1+d_2> m(q-1)$ and in that case we have $\PRM_{d_1+d_2}(m)=\F_q^n$ (see \cite{sorensen}). Therefore, $(C_1\star C_2)^\perp=\{0\}$. In other words, we can only have $(C_1\star C_2)^\perp\neq \{0\}$ if $k_1+k_2\leq n$ for the case of projective Reed-Muller codes. If $k_1+k_2\leq n$, then, according to Theorem \ref{subirhull}, we need to find a vector of Hamming weight $n$ in $(C_1\star C_2)^\perp$. This code is a projective Reed-Muller code if $C_1$ and $C_2$ are projective Reed-Muller codes (see Theorem \ref{dualPRM}), which motivates the following lemma.

\begin{lem}\label{vectorn}
Let $1\leq d < q-2$. Then there is a vector of Hamming weight $n=\frac{q^{m+1}-1}{q-1}$ in $\PRM_d^\perp(m)$.
\end{lem}
\begin{proof}
Let $t$ be a monic polynomial of degree $q-1-d$ such that $t(z)\neq 0$ for every $z\in \fq$. For example, we can choose $t$ as a monic irreducible polynomial in $\fq[x]$. We consider the vector $v=(v_Q)_{Q\in P^m}\in \fq^n$ defined in the following way: if $Q=(0,0,\dots,0,1,z)$, for some $z\in \fq$, we define $v_Q=t(z)$, and $v_Q=1$ otherwise. For $0\leq i\leq q-1-d$, let $t_i$ be the coefficient of $x^i$ in $t$. We consider the following decomposition of $P^m=(\{1\}\times \F_q^m)\cup (\{0\}\times \{1\}\times \F_q^{m-1})\cup \cdots \cup \{(0,0,\dots,0,1)\}=B_m\cup B_{m-1}\cup\cdots \cup B_0$. Let $x_0^{\alpha_0}x_1^{\alpha_1}\cdots x_m^{\alpha_m}=x^\alpha\in \fq[x_0,\dots,x_m]_d$. Then we have
$$
\begin{aligned}
v\cdot \ev(x^\alpha)=\sum_{Q\in B_m}v_Q x^\alpha(Q)+\sum_{Q\in B_{m-1}}v_Q x^\alpha(Q) +\cdots +\sum_{Q\in B_0}v_Q x^\alpha(Q).
\end{aligned}
$$
For $2\leq i\leq m$, we have $v_Q=1$ and 
$$
\sum_{Q\in B_i}v_Q x^\alpha(Q)=\sum_{Q\in B_i} x^\alpha(Q)=0
$$
by Lemma \ref{lemasumafq} and Remark \ref{remafin} because $\alpha_j<q-1$ for $0\leq j \leq m$. 

We study the sum over $B_1$ now. If $\alpha_j=0$ for every $0\leq j \leq m-2$, we have
\begin{equation}\label{ultimasuma}
\sum_{Q=(0,0,\dots,1,z)\in B_1}v_Q x^\alpha(Q)=\sum_{z\in \fq} t(z)x_m^{\alpha_m}(z)=\sum_{l=0}^{q-1-d} t_l \sum_{z\in \fq} x_m^{\alpha_m+l}(z).
\end{equation}
Taking into account that $\alpha_m\leq d$, we have $\alpha_m+l\leq q-1$, and we can only have the equality for some $l\in \{0,\dots,q-1-d\}$ if $\alpha_m=d$. Therefore, by Lemma \ref{lemasumafq}, this sum is equal to 0 unless $\alpha_m=d$, in which case it is equal to $-t_{q-1-d}=-1$ ($t$ is monic). 

On the other hand, if we have $\alpha_j>0$ for some $0\leq j \leq m-2$, then the sum over $B_1$ in (\ref{ultimasuma}) is clearly equal to 0 because all the addends are equal to 0. For the sum over $B_0$, it is clear that this sum is equal to 0 unless $\alpha_m=d$, in which case it is equal to 1. Hence, if $\alpha_m\neq d$, all the sums are equal to 0 and we have $v\cdot \ev(x^\alpha)=0$, and if $\alpha_m=d$, all the sums are 0 besides the sums corresponding to $B_1$ and $B_0$, which are equal to $-1$ and $1$, respectively. Thus, if $\alpha_m=d$ we also  have $v\cdot \ev(x_m^d)=0$. This implies that $v\in \PRM_d^\perp(m)$, and the fact that $v$ has Hamming weight $n$ follows from its definition. 
\end{proof}

Note that, for $d\equiv 0 \bmod q-1$, $d\leq \lfloor m(q-1)/2\rfloor$, we have $(1,\dots,1)$ in $\PRM_d^\perp(m)$ by Theorem \ref{dualPRM}, which is a vector of Hamming weight $n$, and we also know that the corresponding projective Reed-Muller code is contained in its dual \cite[Cor. 10.3]{libroprm}. As a consequence of Theorem \ref{subirhull} and Lemma \ref{vectorn}, we obtain the following.

\begin{prop}\label{propbajarhulleu}
Let $q>2$ and let $1\leq d_1\leq d_2< q-2$. Let $C_i=\PRM_{d_i}(m)$, for $i=1,2$. If $d_1+d_2<q-2$, then, for any $0 \leq \ell\leq \dim C_1$, there is a code $C_{1,\ell}$ monomially equivalent to $C_1$ such that 
$$
\dim C_{1,\ell}\cap C_{2}^\perp =\ell.
$$
\end{prop}
\begin{proof}
We have that $C_1\star C_2=\PRM_{d_1+d_2}(m)$, with $d_1+d_2<q-2$. By Lemma \ref{vectorn}, there is a vector of Hamming weight $n$ in $(C_1\star C_2)^\perp$. Using Theorem \ref{subirhull}, we obtain the result. 
\end{proof}

We can use Proposition \ref{propbajarhulleu} and Theorem \ref{asimetricos} to construct asymmetric and symmetric QECCs. The following result, for the case $m=2$, allows us to vary the parameter $c$ from the codes arising in \cite{sanjoseHullsPRM}, thus giving more flexible parameters.

\begin{thm}\label{reducidoasym}
Let $1\leq d_1\leq d_2 <q-2$ such that $d_1+d_2<q-2$. Then we can construct a quantum code with parameters $[[n,\kappa+c,\delta_z/\delta_x;c]]_q$, for any $0\leq c \leq \dim \PRM_{d_1}(m)$, where $n=\frac{q^{m+1}-1}{q-1}$, $\kappa=n-(\dim \PRM_{d_1}(m)+\dim \PRM_{d_2}(m))$, $\delta_z\geq \wt(\PRM_{d_2}^\perp(m))$ and $\delta_x\geq \wt(\PRM_{d_1}^\perp(m))$.
\end{thm}

The previous result can be used to obtain QECCs that outperform affine Reed-Muller codes, as the next example shows. 

\begin{rem}\label{remcompafin}
For $1\leq d_1\leq d_2 < q-2$ such that $d_1+d_2<q-2$, by Theorem \ref{paramRM} and Theorem \ref{paramPRM} (or directly by counting monomials in $m$ variables of degree $\leq d$ and monomials of degree $d$ in $m+1$ variables) we have $\dim \RM_{d_i}(m)=\sum_{t=0}^{d_i}\binom{t+m-1}{t}=\binom{d_i+m}{m}=\dim \PRM_{d_i}(m)$, for $i=1,2$. Moreover, we have that $\wt(\PRM_{d_i}^\perp(m))=\wt(\PRM_{d_i^\perp}(m))=\wt(\RM_{d_i^\perp -1}(m))=\wt(\RM_{d_i}^\perp(m))$, where $d_i^\perp=m(q-1)-d_i$, for $i=1,2$. Therefore, using Theorem \ref{asimetricos} with $C_i=\RM_{d_i}(m)$ for $i=1,2$, we get an asymmetric QECC with parameters
$$[[q^m,q^m-(\dim \RM_{d_1}(m)+\dim \RM_{d_2}(m)),\wt(\RM_{d_2}^\perp(m))/\wt(\RM_{d_1}^\perp(m))]]_q.
$$
On the other hand, using Theorem \ref{reducidoasym} with $c=0$ we can get an asymmetric QECC with parameters 
$$
[[q^m+\Delta, q^m+\Delta-(\dim \PRM_{d_1}(m)+\dim \PRM_{d_2}(m)),\wt(\PRM_{d_2}^\perp(m))/\wt(\PRM_{d_1}^\perp(m)) ]]_q,
$$
where $\Delta=\frac{q^m-1}{q-1}$. Taking into account the previous discussion, we see that both asymmetric QECCs have the same minimum distances $\delta_z$ and $\delta_x$, but the code obtained using Theorem \ref{reducidoasym} has gained $\Delta$ units in length and dimension, which increases the code rate and decreases the relative minimum distances. If we consider $\frac{\kappa+\delta_z}{n}$ (resp. $\frac{\kappa+\delta_x}{n}$) as a measure of how good a code is in terms of transmission rate and phase-shift error-correction capability (resp. qudit-flip error-correction capability), we see that in this case we obtain asymmetric QECCs with better performance using projective Reed-Muller codes (Theorem \ref{reducidoasym}) than the asymmetric QECCs obtained using affine Reed-Muller codes. 
\end{rem}

The quantum Gilbert-Varshamov bound from \cite{matsumotoimprovedGV} can be used to check the goodness of the parameters of an asymmetric QECC. 

\begin{thm}\label{T:GVmatsumoto}
Assume the existence of integers $n\geq 1$, $1\leq l< n+c$, $\delta_x\geq 1$, $\delta_z\geq 1$, $0\leq c\leq l/2$ such that
$$
\frac{q^{2n-l}-q^{l-2c}}{q^{2n}-1}\left( \sum_{i=0}^{\delta_x-1}\binom{n}{i}(q-1)^i\sum_{j=0}^{\delta_z-1}\binom{n}{j}(q-1)^j-1 \right) <1.
$$
Then there is an EAQECC with parameters $[[n,n-l+c,\delta_z/\delta_x;c]]_q$.
\end{thm}

Given a quantum Gilbert-Varshamov bound, like that one from Theorem \ref{T:GVmatsumoto}, we will say that a code $C$ surpasses it if the bound does not guarantee the existence of a code with the parameters of $C$.

\begin{ex}
Let $q=8$, $m=2$, $d_1=1$ and $d_2=4$. We have $d_1+d_2=5<q-2$, and we can apply Theorem \ref{reducidoasym} to obtain QECCs with parameters $[[73,55+c,6/3;c]]_8$, for $0\leq c\leq \dim \PRM_{1}(2)=3$. All of these codes surpass the Gilbert-Varshamov bound from Theorem \ref{T:GVmatsumoto}. 

\end{ex}
 
In some cases it is possible to obtain nested pairs of codes from subfield subcodes of projective Reed-Muller codes, giving rise to QECCs, as we show in the next result. 

\begin{prop}\label{P:eusscPRM}
Let $d_1,d_2$ such that $d_1+d_2=\lambda (q^s-1)$, with $1\leq \lambda \leq m$. Then we have $(\PRM_{d_1}(q^s,m))_q\subset ((\PRM_{d_2}(q^s,m))_q)^\perp$, and we can construct a QECC with parameters $[[n,\kappa,\delta_z/\delta_x]]_q$, where $n=\frac{q^{s(m+1)}-1}{q^s-1}$, $\kappa=n-\dim (\PRM_{d_1}(q^s,m))_q-\dim (\PRM_{d_2}(q^s,m))_q$, $\delta_z\geq \wt(((\PRM_{d_2}(q^{s},m))_{q})^{\perp})$ and $\delta_x\geq \wt(((\PRM_{d_1}(q^{s},m))_{q})^{\perp})$.
\end{prop}
\begin{proof}
By hypothesis we have $d_1\equiv d_2^\perp \bmod q^s-1$ and $d_1\leq d_2^\perp$. By \cite[Lem. 10.7]{libroprm}, $\PRM_{d_1}(q^s,m)\subset\PRM_{d_2}^\perp(q^s,m)$. This implies
$$
(\PRM_{d_1}(q^s,m))_q \subset (\PRM_{d_2}^\perp(q^s,m))_q\subset ((\PRM_{d_2}(q^s,m))_q)^\perp.
$$
The parameters follow from Theorem \ref{asimetricos}.
\end{proof}

\begin{rem}\label{R:degenerate}
In general, we do not know $\wt(((\PRM_{d_i}(q^{s},m))_{q})^{\perp})$, for $i=1,2$. Nevertheless, we can compute this minimum distance with Magma \cite{magma}. In some cases, the code $(\PRM_{d_i}(q^{s},m))_{q}$ can be degenerate, that is, its generator matrix has a column of zeroes. In that case, the minimum distance of its dual is equal to 1. This problem can easily be avoided by considering the results from \cite{sanjoseRecursivePRM}, where some particular degrees are considered such that $(\PRM_{d_i}(q^{s},m))_{q}$ is nondegenerate. For those degrees, the resulting codes have good parameters and we have formulas for the dimension of the subfield subcode (for the case $m=2$, one can also use the results from \cite{sanjoseSSCPRM} for the dimension). In the next example, we show how to use those results together with Proposition \ref{P:eusscPRM}.
\end{rem}

\begin{ex}
Let $m=2=s$. From \cite{sanjoseRecursivePRM} (or \cite{sanjoseSSCPRM}), if $d_i\equiv 0 \bmod (q^s-1)/(q-1)$, then $(\PRM_{d_i}(q^{s},2))_{q}$ is non degenerate, for $i=1,2$. If we set $q=2$, from Proposition \ref{P:eusscPRM} we obtain a code with parameters $[[73,19,9/9]]_2$ with $d_1=d_2=7$. For $q=3$, we obtain the parameters $[[91,73,4/4]]_3$ with $d_1=d_2=4$, and $[[91,12,36/4]]_2$ with $d_1=4$, $d_2=12$. All of them surpass the Gilbert-Varshamov bound from Theorem \ref{T:GVmatsumoto}.
\end{ex}

\section{Hermitian construction}\label{secherm}
In the Hermitian case, we can also obtain quantum codes with projective Reed-Muller codes without entanglement assistance. Recall that in this case we have to consider codes over $\F_{q^2}$ and the Hermitian product. The following result from \cite[Thm. 2.1]{haochenhull} is analogous in part to Theorem \ref{subirhull} for the Hermitian product.

\begin{thm}\label{T:hao}
Let $C\subset \F_{q^2}$ be a linear code. If there is a vector $v\in ((C\star C^q)^\perp)_q$ with $\wt(v)=n$, then $\langle v \rangle \star C\subset (\langle v \rangle \star C)^{\perp_h}$, i.e., $\langle v \rangle \star C$ is self-orthogonal with respect to the Hermitian product.
\end{thm}
Note that, unlike Theorem \ref{subirhull}, the previous result does not cover all the possible values of the dimension of the Hermitian hull, we go directly to the case $\Hull_H(C)=C$. The rest of the values are covered by the following result from \cite{grasslvariableentanglement}.

\begin{thm}\label{T:subirhullhermgrassl}
Let $q>2$ and let $C\subset \F_{q^2}^n$ with $\dim \Hull_H(C)=\ell$. Then there exists a monomially equivalent code $C_{\ell'}$ with $\dim \Hull_H(C_{\ell'})=\ell'$, for each $0\leq \ell' \leq \ell$. 
\end{thm}

Therefore, using Theorem \ref{T:hao} and Theorem \ref{T:subirhullhermgrassl}, if we find a vector $v\in ((C\star C^q)^\perp)_q$ with $\wt(v)=n$, then we can find an equivalent code $C_{\ell}$ such that $\dim \Hull_H(C_{\ell})=\ell$, for $0\leq \ell \leq \dim C$, if $q>2$. 

With respect to projective Reed-Muller codes, we start by determining some of these codes which are self-orthogonal with respect to the Hermitian product, without considering equivalent codes.

\begin{prop}\label{contencionherm}
Let $d=\lambda (q-1)$ with $1\leq \lambda \leq m$. Then $\PRM_d(q^2,m)\subset \PRM_d^{\perp_h}(q^2,m)$. As a consequence, if $d\equiv 0\bmod q-1$ and $d\not \equiv 0\bmod q^2-1$, we can construct a QECC with parameters $[[n,\kappa,\delta]]_q$, where $n=\frac{q^{2(m+1)}-1}{q^2-1}$, $\kappa=n-2(\dim \PRM_d(q^2,m))$ and $\delta \geq \wt(\PRM_{d}^\perp(q^2,m))$.  
\end{prop}
\begin{proof}
We have
$$
\begin{aligned}
((\PRM_d(q^2,m) \star \PRM_d(q^2,m)^q)^\perp)_q \supset (\PRM_{d+qd}^\perp(q^2,m))_q.
\end{aligned}
$$
By Theorem \ref{dualPRM}, if we have $d(q+1)\equiv 0 \bmod q^2-1$, with $d(q+1)\leq m(q^2-1)$, then $(1,\dots,1)\in (\PRM_{d+qd}^\perp(q^2,m))_q$ and we can apply Theorem \ref{T:hao} with $C=\PRM_d(q^2,m)$. The parameters of the quantum code are deduced from Theorems \ref{paramPRM}, \ref{dualPRM} and \ref{hermitica}.
\end{proof}

With this construction (and, in general, with the results of this section) we manage to construct very long codes over small finite field sizes. To check the performance of these codes, we introduce now another quantum Gilbert-Varshamov bound from \cite{quantumGVexigente}, which seems to be more difficult to surpass for symmetric QECCs than the bound from Theorem \ref{T:GVmatsumoto}.

\begin{thm}\label{T:GVHermitica}
Suppose that $n>\kappa$, $\delta\geq 2$ and $n\equiv \kappa \bmod 2$. Then there exists a pure stabilizer quantum code $[[n,\kappa,\delta]]_q$, provided that
$$
\frac{q^{n-\kappa+2}-1}{q^2-1}>\sum_{i=1}^{\delta-1}(q^2-1)^{i-1}\binom{n}{i}.
$$
\end{thm}

\begin{ex}
Let $q=2$ and $m=3$. For $d=q-1=1$ we can apply Proposition \ref{contencionherm} to obtain a QECC with parameters $[[85,77,3]]_2$, which is optimal according to \cite{codetables}. For $m=4$, we get a QECC with parameters $[[341,331,3]]_2$. 

Consider now $q=3$ and $d=q-1=2$. For $m=2$, we obtain the parameters $[[91,79,4]]_3$, and for $m=3$ we obtain $[[820,800,4]]_3$. All the codes in this example surpass the quantum Gilbert-Varshamov bound from Theorem \ref{T:GVHermitica}. 

\end{ex}

When $d\not\equiv 0 \bmod q-1$, we can use Theorems \ref{T:hao} and \ref{T:subirhullhermgrassl} to construct quantum codes with a variable amount of entanglement using equivalent codes in some cases. The idea of the proof of the following result can be regarded as an extension of the proof from \cite[Thm. 4]{ball} for projective Reed-Solomon codes.

\begin{thm}\label{bajarcherm}
Let $1\leq d < q-2$. Then we can construct an EAQECC with parameters $[[n,\kappa+c,\delta;c]]_{q}$, for any $0\leq c \leq \dim \PRM_d(q^2,m)$, where $n=\frac{q^{2(m+1)}-1}{q^2-1}$, $\kappa=n-2(\dim \PRM_d(q^2,m))$ and $\delta \geq \wt(\PRM_{d^\perp}(q^2,m))$.  
\end{thm}
\begin{proof}
We use Theorems \ref{T:hao} and \ref{T:subirhullhermgrassl} with $C=\PRM_d(q^2,m)$. First, we find $w\in ((C\star C^q)^\perp)_q $ with $\wt(v)=n$. Let $t$ be a monic polynomial with coefficients in $\F_{q^2}$ of degree $q-1-d$ such that $t(z)\neq 0$ for every $z\in \F_{q^2}$, and $v\in \F_{q^2}^n$ as in Lemma \ref{vectorn}. Then we consider $w:=v^{q+1}$, where the power is taken component wise. Clearly $w\in \F_q^n$, $\wt(w)=n$, and we will prove that $w\in (C\star C^q)^\perp$. Let $x^\alpha,x^\beta\in \F_{q^2}[x_0,\dots,x_m]_d$. Using the decomposition $P^m=B_m\cup B_{m-1}\cup \cdots \cup B_0$ from the proof of Lemma \ref{vectorn} we have that
$$
w \cdot (\ev(x^\alpha)\star \ev(x^{q\beta})) =\sum_{i=0}^m \sum_{Q\in B_i} v_Q^{q+1} x^{\alpha+q\beta}(Q).
$$
For $2\leq i \leq m$, we have $v_Q=1$ and
$$
\sum_{Q\in B_i}v_Q^{q+1} x^{\alpha+q\beta}(Q)=\sum_{Q\in B_i} x^{\alpha+q\beta}(Q)=0
$$
because $\alpha_j+q\beta_j<q-1+q(q-1)=q^2-1$ for $0\leq j \leq m$.

On the other hand, if $\alpha_j=0$ for $0\leq j \leq m-2$, we have
\begin{equation}\label{E:sumaherm}
\sum_{Q=(0,0,\dots,1,z)\in B_1}v_Q^{q+1} x^{\alpha+q\beta}(Q)=\sum_{z\in \F_{q^2}} t(z)^{q+1}z^{\alpha_m+q\beta_m}=\sum_{l=0}^{(q+1)(q-1-d)} r_l \sum_{z\in \F_{q^2}} z^{l+\alpha_m+q\beta_m},
\end{equation}
where $r_l$ is the coefficient of $x^l$ in $t^{q+1}$. Taking into account that $l+\alpha_m+q\beta_m\leq (q+1)(q-1-d)+d+qd=q^2-1$, we see that this sum is equal to 0 unless $\alpha_m=\beta_m=d$, in which case it is equal to $-r_{(q+1)(q-1-d)}=-1$ (since $t$ is monic), because of Lemma \ref{lemasumafq} and Remark \ref{remafin}. If we have $\alpha_j>0$ for some $0\leq j \leq m-2$, then all the addends in (\ref{E:sumaherm}) are equal to 0. For the sum over $B_0$, it is clear that this sum is equal to 0 unless $\alpha_m=\beta_m=d$, in which case it is equal to 1.

Therefore, if $\alpha_m\neq d$ or $\beta_m\neq d$, all the sums are equal to 0 and we have $w \cdot (\ev(x^\alpha)\star \ev(x^{q\beta}))=0$, and if $\alpha_m=\beta_m=d$, all the sums are 0 besides the sums corresponding to $B_1$ and $B_0$, which are equal to $-1$ and $1$, respectively. Thus, if $\alpha_m=\beta_m=d$ we also  have $w \cdot (\ev(x^\alpha)\star \ev(x^{q\beta}))=0$. This implies that $w\in (C\star C^q)^\perp$. Therefore, $\langle w\rangle \star C\subset (\langle w\rangle \star C)^{\perp_h}$ by Theorem \ref{T:hao}, and we finish the proof by applying the Hermitian construction from Theorem \ref{hermitica} to $\langle w \rangle \star C$ and considering Theorem \ref{T:subirhullhermgrassl}.
\end{proof}

The argument in Remark \ref{remcompafin}, changing $q$ with $q^2$ and $d_1=d_2=d$ (note that $d<q-2$  implies $2d<q^2-2$), shows that, using the QECCs from Theorem \ref{bajarcherm}, we can obtain QECCs with $\frac{q^{2m}-1}{q^2-1}$ extra length and dimension with respect to the affine case. In the next example, we show some codes obtained from Theorem \ref{bajarcherm} with good parameters. 

\begin{ex}
We consider $q=4$ and $m=2$. Therefore, we work over the field $\F_{4^2}$ and we obtain codes of length $\frac{q^{2(m+1)}-1}{q^2-1}=273$. For $d=1$ we can apply Theorem \ref{bajarcherm} to obtain a QECC with parameters $[[273,267,3]]_4$. This code improves the parameters of the code $[[273,265,3]]_4$ from \cite{quantumtwisted} and surpasses the quantum Gilbert-Varshamov from Theorem \ref{T:GVHermitica}.

For $q=5$, $m=2$ and $d=1,2,$ we obtain the parameters $[[651,645,3]]_5$ and $[[651,639,4]]_5$, respectively. The first one improves the parameters $[[651,642,3]]_5$ and $[[652,644,3]]_5$ obtained in \cite{quantumtwisted}, and the second one improves the parameters $[[651,636,4]]_5$ and $[[652,638,4]]_5$ from \cite{quantumtwisted}. Both of them exceed the quantum Gilbert-Varshamov bound from Theorem \ref{T:GVHermitica}.
\end{ex}

Similarly to the previous section, we can also use subfield subcodes of projective Reed-Muller codes in some cases. For the Hermitian product, we consider projective Reed-Muller codes over $\F_{q^{2s}}$ such that their subfield subcodes with respect to the extension $\F_{q^{2s}}/\F_{q^2}$ are self-orthogonal with respect to the Hermitian product, which gives rise to QECCs over $\F_q$ using the Hermitian construction from Theorem \ref{hermitica}. 

\begin{prop}\label{P:subfield}
Let $d=\lambda (q^{2s}-1)/(q+1)$ with $1\leq \lambda \leq m$. Then $(\PRM_d(q^{2s},m))_{q^2}\subset ((\PRM_d(q^{2s},m))_{q^2})^{\perp_h}$. As a consequence, we can construct a QECC with parameters $[[n,\kappa,\delta]]_q$, where 
$$n=\frac{q^{2s(m+1)}-1}{q^{2s}-1},\;\kappa=n-2(\dim (\PRM_d(q^{2s},m))_{q^2}) \text{ and } \delta\geq \wt(((\PRM_d(q^{2s},m))_{q^2})^\perp).$$ 
\end{prop}
\begin{proof}
We have
$$
\begin{aligned}
(((\PRM_d(q^{2s},m))_{q^2} \star ((\PRM_d(q^{2s},m))_{q^2})^q)^\perp)_q &\supset ((\PRM_d(q^{2s},m) \star (\PRM_d(q^{2s},m))^q)^\perp)_q \\
&\supset (\PRM_{d+qd}^\perp(q^{2s},m))_q.
\end{aligned}
$$
The rest of the proof follows as in the proof of Proposition \ref{contencionherm}, changing $q^2$ with $q^{2s}$. 
\end{proof}

As we stated in Remark \ref{R:degenerate}, we can use the results from \cite{sanjoseRecursivePRM} to obtain QECCs with good parameters, whose minimum distance can be computed using Magma \cite{magma}. We show this in the next example.

\begin{ex}
By \cite[Cor. 2 and Cor. 3]{sanjoseRecursivePRM}, if $d\equiv 0 \bmod (q^{2s}-1)/(q^2-1)$, then we have recursive formulas for the dimension of $(\PRM_d(q^{2s},m))_{q^2}$, and we know that $(\PRM_d(q^{2s},m))_{q^2}$ is non degenerate (this can also be seen recursively, using what we know for the case $m=1$ from \cite{sanjoseSSCPRS}). 

Let $q=2$ and $m\geq 2$. Then $(q^{2s}-1)/(q+1)=(q^{2s}-1)/(q^2-1)=(4^s-1)/3$. Hence, by Proposition \ref{P:subfield}, if we consider $d=\lambda (4^s-1)/3$, for some $1\leq \lambda \leq m$, then we have $(\PRM_d(4^{s},m))_{q^2}\subset ((\PRM_d(4^{s},m))_{q^2})^{\perp_h}$, and by the previous discussion $((\PRM_d(4^{s},m))_{q^2})^{\perp_h}$ is non degenerate. For example, for $m=s=2$, we can consider $\lambda=1$ and $d=5$. The corresponding quantum code has parameters $[[273, 255, 4]]_2$, which improves the parameters $[[274,248,4]]_2$ and $[[273,246,4]]_2$ from \cite{quantumtwisted}. If we consider $m=3$ instead, we obtain the parameters $[[4369, 4337,4]]_2$. Both of the codes surpass the Gilbert-Varshamov bound from Theorem \ref{T:GVHermitica}.
\end{ex}

\begin{rem}
If $q>2$, one can also use Theorem \ref{T:subirhullhermgrassl} with Proposition \ref{contencionherm} or Proposition \ref{P:subfield} to obtain EAQECCs with a variable amount of entanglement. 
\end{rem}

\bibliographystyle{abbrv}
%\bibliography{BIBR}

\end{document}